# Wave-Packet Surface Propagation for Light-Induced Molecular Dissociation

**Shengzhe Pan[1], Zhaohan Zhang[2], Chenxi Hu[2], Peifen Lu[1], Xiaochun Gong[1], Ruolin Gong[1], Wenbin Zhang[1], Lianrong Zhou[1], Chenxu Lu[1], Menghang Shi[1], Zhejun Jiang[1], Hongcheng Ni[1], Feng He[2,3], Jian Wu[1,3]\***

[1]*State Key Laboratory of Precision Spectroscopy, East China Normal University, Shanghai 200241, China*

[2]*Key Laboratory for Laser Plasmas (Ministry of Education) and School of Physics and Astronomy, Collaborative Innovation Center of IFSA (CICIFSA), Shanghai Jiao Tong University, Shanghai 200240, China*

[3]*CAS Center for Excellence in Ultra-intense Laser Science, Shanghai 201800, China*

**Abstract**

Recent advances in laser technology have enabled tremendous progress in photochemistry, at the heart of which is the breaking and formation of chemical bonds. Such progress has been greatly facilitated by the development of accurate quantum-mechanical simulation method, which, however, does not necessarily accompany clear dynamical scenarios and is rather often a black box, other than being computationally heavy. Here, we develop a wave-packet surface propagation (WASP) approach to describe the molecular bond-breaking dynamics from a hybrid quantum-classical perspective. Via the introduction of quantum elements including state transitions and phase accumulations to the Newtonian propagation of the nuclear wave-packet, the WASP approach naturally comes with intuitive physical scenarios and accuracies. It is carefully benchmarked with the $H_2^+$ molecule and is shown to be capable of precisely reproducing experimental observations. The WASP method is promising for the intuitive visualization of strong-field molecular dynamics and is straightforwardly extensible toward complex molecules.

\* jwu@phy.ecnu.edu.cn

# Introduction

Coherent control of molecular breakup and formation has been a heated topic of discussion lately thanks to recent advances in laser technology that make tailored intense laser pulses routinely available [1-5]. The ingeniously sculptured laser fields carry specific frequencies tailored to populate certain states and initiate controlled or directed chemical reactions [6-15]. The molecular dynamics during chemical reactions can be, in principle, well described by quantum mechanics via the solution of the time-dependent Schrödinger equation (TDSE). However, even for the simplest neutral molecule of $H_2$, an *ab initio* numerical simulation is still challenging due to the large number of degrees of freedom involved in a chemical reaction. Furthermore, an intuitive physical picture remains evanescent even if an agreement with experimental results can be reached. Therefore, a general approach with low computational cost and a clear physical picture is highly desirable for describing molecular dynamical reactions under intense laser fields.

In this article, we develop an intuitive WAve-packet Surface Propagation (WASP) approach for the description of the molecular dissociation process. It is realized via the introduction of quantum elements including state transitions and phase accumulations into the classical propagation of nuclear wave-packets (NWPs) on the potential energy surfaces of a stretching molecule. Using a classical description of the nuclear dynamics substantially alleviates the computational overhead, while keeping quantum elements during transitions preserves the essential physics crucial to light-induced chemical reactions. The WASP approach is carefully benchmarked against the $H_2^+$ molecule and is shown to be capable of precisely modeling experimental observations. In particular, the WASP model reproduces the feature-rich two-dimensional (2D) momentum distributions and the laser-phase-dependent directional ejection of the protons in the dissociative single ionization of $H_2$ driven by an orthogonal two-color (OTC) laser pulse. The present WASP approach provides a hybrid quantum-classical perspective to precisely simulate the strong-field molecular dissociation, which is both highly efficient and physically transparent in comparison to *ab initio* quantum simulation methods.

## Results and Discussion

**Construction of dissociating NWP**

The essence of the WASP model is the construction of the dissociating NWP by including the state transitions and phase accumulations to its Newtonian propagation along various pathways as the molecule stretches. For the sake of clarity and simplicity, we restrict our study to 2D and demonstrate the light-induced molecular dissociation dynamics using the simplest molecule $H_2^+$ exposed to an OTC laser pulse, as schematically illustrated in Fig. 1. The OTC laser pulse is expressed as $\boldsymbol{E}(t) = E_{FW} \exp(-2\ln 2\, t^2/\tau_{FW}^2) \cos(\omega_{FW} t + \phi_{CEP}) \boldsymbol{e_y} + E_{SH} \exp(-2\ln 2\, t^2/\tau_{SH}^2) \cos(\omega_{SH} t + 2\phi_{CEP} + \phi_L) \boldsymbol{e_z}$. The fundamental-wave (FW) pulse is polarized along the $y$ axis and the second-harmonic (SH) pulse is polarized along the $z$ axis. Here $\phi_{CEP}$ and $\phi_L$ denote the carrier-envelope phase of the FW pulse and the relative phase between the two waves of the OTC field, respectively. The geometry of the present work is thereby set up such that the molecule resides in the $y$-$z$ plane where the momentum distribution of ejected protons is collected and the potential energy surfaces of the homonuclear diatomic molecule is reduced to the potential energy curves with a single coordinate for nuclear motions. The initial NWP is launched at the equilibrium internuclear distance of the neutral $H_2$ molecule upon electron removal at the peak of the OTC pulse. The subsequent propagation of the NWP on the potential energy curves of the $H_2^+$ cation is described as the Newtonian motion of a classical particle (see Methods), which has a certain probability to absorb or emit photons, enabling the NWP to transit among different electronic states and dissociate along various pathways. During the dissociation of $H_2^+$, only the two lowest electronic states ($1s\sigma_g$ and $2p\sigma_u$) are considered. The time-dependent internuclear distance of the NWP for pathway $J$ is denoted as $R_J(t)$ with the bond stretching time $t$. The overall transition amplitude $A_J$ depends on the field strength at each state-transition internuclear distance (details are discussed in the following section). The dissociating NWPs of different pathways with the same kinetic energy release (KER) would interfere with each other, with individual phases $\varphi_J$ accumulated along the respective dissociation pathways (see Methods), which leads to the rich structure of the observed KER spectrum on the detector, as illustrated in the right panel of Fig. 1.

To construct the 2D momentum distribution of the dissociating NWP versus the two-color laser phase $\phi_L$, the pathway-resolved dissociation probabilities at various molecular orientations $\theta_p$ in the polarization plane are calculated using the WASP model. The rotation of the molecular axis occurs on a time scale much longer than that of dissociation and thus is not considered here. The resulted 2D NWP superimposed from various dissociation pathways can then be written as

$$\psi(E_k, \theta_p, \phi_L) = \sum_J \psi_J(E_k, \theta_p, \phi_L) = \sum_J [G_J(E_k) A_J(\theta_p, \phi_L) e^{-i\varphi_J}], \quad (1)$$

where $\psi_J$ is the NWP for pathway $J$ and $G_J$ is the kinetic energy distribution formulated as a Gaussian distribution with the width depending on the involved photon number and pulse duration. The 2D NWP constructed above is subsequently employed to extract the physical quantities measurable in experiments, such as the momentum and KER distributions of the ejected protons.

**Nonlinear transition amplitudes**

Extending the present model to cover the multiphoton regime driven by strong laser fields, we include the nonlinear multiphoton transitions beyond the perturbation theory [16-19]. To maintain the consistency between the one-photon and multiphoton transitions, we deduce the transition amplitude from a unified Rabi model. For the two lowest potential energy curves of $H_2^+$, the one-photon transition amplitude at the internuclear distance where the potential energy gap between two involved electronic states matching the photon energy is formulated as (see Methods)

$$A_{1\omega} = \Omega_{1\omega} \tau_{1\omega}, \quad (2)$$

where $\tau_{1\omega}$ denotes the effective duration of the one-photon transition in molecular systems and $\Omega_{1\omega} = D(R_{1\omega})\mathcal{E}(t)/2$ denotes the dipole interaction at the resonant internuclear distance $R_{1\omega}$ with $D$ representing the dipole moment between the two electronic states and $\mathcal{E}$ standing for the strength of the external field in the molecular frame.

Analogous to the one-photon transition amplitude, the multiphoton transition amplitude can also be deduced using the multiphoton Rabi model, where the *n*-photon coupling strength is formulated as [20]

$$\Omega_n = \Omega_0 \prod_{k=2}^{n} \left(\frac{\Omega_0}{\Delta_k}\right). \tag{3}$$

Here $\Omega_0$ and $\Delta_k$ denote the coupling and intermediate detuning at each transition (see Supplementary Materials [21] for details). The overall multiphoton transition amplitude, e.g., the transition amplitude of the net-two-photon dissociation pathway of $H_2^+$, can be deduced by combining the three-photon absorption at $R_{3\omega}$ and the one-photon emission at $R_{1\omega}$, resulting in

$$A_{2\omega} = \frac{\Omega_{3\omega}}{2\omega}\frac{\Omega_{3\omega}}{2\omega}\Omega_{3\omega}\frac{\Omega_{1\omega}}{\Delta V_{u(31)}}\tau_{2\omega}. \tag{4}$$

Here $\Omega_{3\omega} = D(R_{3\omega})\mathcal{E}(t)/2$ denotes the dipole interaction at the three-photon resonant internuclear distance $R_{3\omega}$, $\tau_{2\omega}$ denotes the effective duration of the net-two-photon dissociation pathway, and the intermediate detuning of the subsequent one-photon emission is given by $\Delta V_{u(31)} = V_u(R_{3\omega}) - V_u(R_{1\omega})$ with $V_u$ the potential energy curve of the 2*p*σ*u* state (see Supplementary Materials [21] for details). The nonlinear transition amplitude has been benchmarked and confirmed in experiments using a single near-infrared laser pulse (see Supplementary Materials [21] for details), where the yield ratio of the net-two-photon pathway to the one-photon pathway has been utilized to calibrate the laser intensity of the interaction region [22].

Other than the *n*-photon resonant transition, another strong laser-molecule coupling is present during the molecular dissociation, which we name as dynamical Rabi coupling, where the electron continuously hops between two electronic states governed by the area theorem driven by the laser pulse [23-25]. The off-resonant transition internuclear distance $R_{\mathcal{R}_n}$ of the dynamic Rabi coupling can be obtained based on the area theorem, i.e., $\int \omega_r(R_{\mathcal{R}_n}, t)\mathrm{d}t = n\pi$, where $\omega_r(R, t) =$ $\sqrt{[\Delta E(R,t) - \hbar\omega]^2 + [D(R,t)\mathcal{E}(t)]^2}$ denotes the coupling frequency and $n$ denotes the sequence number of transitions. Consequently, the transition amplitude of the one-photon-one-Rabi-coupling pathway can be described as

$$A_{1\omega-1\mathcal{R}} = \Omega_{1\omega} \frac{\Omega_{\mathcal{R}_1}}{\Delta V_{g(11')}} \frac{\Omega_{\mathcal{R}_2}}{\Delta V_{u(12')}} \tau_{1\omega-1\mathcal{R}}, \qquad (5)$$

where the intermediate detunings of the subsequent off-resonant transitions are defined as $\Delta V_{g(11')} = -V_g(R_{1\omega}) + V_g(R_{\mathcal{R}_1})$ and $\Delta V_{u(12')} = V_u(R_{1\omega}) - V_u(R_{\mathcal{R}_2})$ (see Supplementary Materials [21] for details).

**Anisotropic proton spectra**

To evaluate the accuracy of a numerical method, we compare the simulation results with the experimental observations. Experimentally, a cold-target recoil ion momentum spectroscopy reaction microscope [26,27] is employed to measure the momenta of the proton from the dissociation of $H_2^+$ driven by an OTC femtosecond laser pulse (see Methods). Figure 2A shows the measured 2D momentum distribution of the ejected protons integrated over all the two-color relative phase $\phi_L$ with the OTC polarization as shown in the inset. Here the peak intensity of the SH pulse is $I_{SH} = 4 \times 10^{13}$ W/cm$^2$ and the peak intensity of the FW pulse is $I_{FW} = 1 \times 10^{12}$ W/cm$^2$. The numerical result is obtained by the present WASP model with the same OTC pulse, as shown in Fig. 2B, which agrees very well with experimental observations. In the simulation, we have scanned the carrier-envelop phase $\phi_{CEP}$ and two-color relative phase $\phi_L$ of the laser pulses, and the 2D momentum distribution is averaged over $\phi_{CEP}$ and $\phi_L$ to reach a consensus with the experimental measurement without the carrier-envelope phase-locking.

With the intuitive physical picture offered by the present WASP approach, we can infer relevant dissociation pathways involved in the molecular dissociation process. The high-energy protons with a momentum of 10.7 a.u. (KER ~ 1.7 eV) ejecting along the $y$ axis (0°) in Figs. 2A and 2B are mainly produced from the dissociation of the $1\omega_{SH}$ pathway (the NWP moving outwards on the $1s\sigma_g$ curve, followed by absorbing one photon of $\omega_{SH}$ and dissociating along the $2p\sigma_u$ curve). The low-energy protons with a momentum of 7.3 a.u. (KER ~ 0.8 eV) ejecting along the $z$ axis (90°) in Fig. 2B are attributed to the $1\omega_{FW}$ pathway (the NWP moving outwards on the $1s\sigma_g$ curve, followed by absorbing one photon of $\omega_{FW}$ and dissociating along the $2p\sigma_u$ curve). The main reason for the low contrast ratio of the $1\omega_{FW}$ pathway in experiments is the relatively low efficiency of the detector.

**Signatures of the butterfly structure**

The most prominent feature of the proton momentum distribution in Fig. 2A is the butterfly structure at very low energy (KER < 0.4 eV with momenta less than 5.2 a.u.) originating from the conjunct driving of the orthogonally polarized FW and SH fields, which is verified by individually switching off the FW or SH pulse in the WASP simulation (see Supplementary Materials [21] for details). It is mainly produced via the $1\omega_{SH} - 1\omega_{FW}$ pathway, i.e., the NWP propagating outwards on the $1s\sigma_g$ curve firstly transits to and propagates along the $2p\sigma_u$ state by absorbing one photon of $\omega_{SH}$, subsequently de-excites back to the $1s\sigma_g$ state by emitting one photon of $\omega_{FW}$ and eventually dissociate along the $1s\sigma_g$ curve. For the OTC laser field, the protons produced in the $1\omega_{SH} - 1\omega_{FW}$ dissociation pathway mainly eject along 45° due to the equal intensity dependence on the two orthogonal laser fields. However, as shown in the zoom-in in Fig. 2C, the butterfly structure is not strictly along 45° but slightly bends with a trend towards $y$ axis with decreasing momenta, which is attributed to stronger couplings with the $y$-polarized SH field (whose intensity is stronger than the $z$-polarized FW field, thus inducing more nonlinear couplings). This effect is modeled by the dynamical Rabi couplings in our WASP simulation. Figure 2C shows a schematic illustration of a typical two-color dissociation pathway with the dynamical Rabi coupling, named $1\omega_{SH} + 1\mathcal{R}_{SH} - 1\omega_{FW}$ pathway. After the NWP absorbs one photon of $\omega_{SH}$, it may hop two times between two electronic states, resulting in one up-down cycle of dynamical Rabi couplings driven by the SH field, followed by the emission of one photon of $\omega_{FW}$ at a larger internuclear distance. An increase in the number of times of dynamical Rabi couplings leads to a decrease in the KER of the dissociative fragments and an increased intensity dependence on the laser field [25]. By switching off the dynamical Rabi couplings in the WASP simulation, as shown in Fig. 2D, the proton yield at very low kinetic energy dramatically decreases, and the above-observed butterfly structure becomes strictly along 45°. Thus, the protons in the near-zero KER region are produced by the strong dynamic Rabi couplings of the SH field, which drives the up-down hopping of the NWP many times between two electronic states.

The 2D momentum distribution of the protons shown here is also benchmarked against the full-quantum simulation by numerically solving the TDSE (see Supplementary Materials [21] for details). In principle, the TDSE calculation, which covers all degrees of freedom including molecular vibration and rotation, electron excitation and ionization should give precise results. However, it is challenging to cover all degrees of freedom if strong near-infrared laser pulses are used. For complex molecules, the computation complexity is increased exponentially. Therefore, the Frank-Condon and Bohr-Oppenheimer approximations are usually adopted. For the TDSE calculations in the current work, the simulation box is restricted by taking into account the near-threshold low-energy protons and high-energy outgoing protons simultaneously, and the electron dynamics is restricted between the $1s\sigma_g$ and $2p\sigma_u$ states. Based on these constraints, the WASP method can achieve better agreement with the experimental observations compared to the TDSE method.

**Laser-phase-dependent directional proton emission**

Directional breaking of the molecular bonds is essential for the stereochemical reaction, providing a straightforward route to control the fate of molecules. It can be achieved by coherently superimposing outgoing NWPs of opposite parities in light-induced molecular dissociation, which strongly depends on the laser phase and the KER of the ejected protons. To clearly demonstrate the directional bond breaking, we slightly adjust the laser intensities to be $I_{SH} = 6 \times 10^{12}$ W/cm$^2$ and $I_{FW} = 8 \times 10^{13}$ W/cm$^2$ so that more dissociation pathways with opposite parities participate with appropriate yields. The degree of the directional bond breaking is quantified by calculating the normalized differential yield of the protons emitting in one direction at a specific two-color phase of $\phi_L$ (see Methods). The measured and WASP simulated asymmetric distributions of the ejected protons as a function of the relative phase $\phi_L$ and the KER are shown in Figs. 3A and 3B. Here, we select the molecular orientation of $30° < \theta_p < 60°$ to demonstrate the asymmetric distributions at different KER regions. Clearly asymmetric structures are visible for the low-energy and high-energy regions as a function of the relative phase $\phi_L$. The corresponding laser-phase-dependent asymmetries integrated over the regions of 0.4 eV < KER < 0.6 eV and 1.2 eV < KER < 1.4 eV are shown in the top panels. The phase shift of $\sim \pi/2$ is clearly observed from the two oscillatory curves. The main asymmetric distribution and the phase shift in simulated results are well in line with the experimental measurements.

To further visualize the underlying physics and resolve the interaction of various dissociation pathways, we show the simulated KER spectrum as a function of the proton emission direction $\theta_p$ in Fig. 4A, where the directions of $\theta_p = 0°$ and 90° are along the polarization axis of the SH and FW pulses, respectively. Figure 4B shows the differential KER spectra calculated by subtracting the simulated 2D KER spectra including the phase accumulation in Fig. 4A from the one without the phase accumulation. The differential spectra peaked around 0.5 eV and 1.2 eV reveal the destructive interference of two NWPs with opposite parities. The KER spectrum of protons emitting to $\theta_p = 45°$ is plotted as an example in Fig. 4C as gray shaded areas. Here, the solid and dashed curves denote the NWPs eventually dissociating from the $2p\sigma_u$ and $1s\sigma_g$ states, respectively. Two KER ranges are clearly identified where NWPs of opposite parities overlap for the directional bond breaking: one around 0.5 eV corresponding to the interference of the $1\omega_{FW}$ and $1\omega_{SH} - 1\omega_{FW}$ pathways, and the other around 1.2 eV corresponding to the interference of the $3\omega_{FW} - 1\omega_{SH}$ and $3\omega_{FW} - 1\omega_{SH} + 1\omega_{FW}$ pathways. On the other hand, although several NWPs of different dissociation pathways overlap at the very low KER region below 0.4 eV, no directional bond breaking versus the laser phase is observed since these NWPs have the same parity. Obviously, the WASP model captures the essential experimental observations and provides a physically transparent route to uncover quantum interferences in light-induced molecular dissociation.

## Discussion

In summary, we have developed a WASP approach to simulate strong-field molecular dissociation from a hybrid quantum-classical perspective. By introducing state transitions and phase accumulations to the classical propagation of the NWP on the involved potential energy surfaces of the dissociating molecule, the WASP approach combines the advantages of the clarity of a classical description and the precision of quantum evolution. The WASP model has been benchmarked against the simplest molecule of $H_2^+$ and was shown to accurately predict the ionic momentum distribution, as verified by our experiment and full-quantum simulations. In particular, the WASP model includes the dynamical Rabi coupling induced up-down hopping of the NWP and is capable of simulating the very low energy proton emission, e.g., the butterfly structure produced by an OTC laser pulse, in line with experimental observations. The WASP approach features a low computational cost and an intuitive physical picture while at the same time capturing the essential quantum mechanical dynamical evolution of the molecular dissociation process. We note that the flexibility of the present WASP approach comes at the price of the absence of certain quantum effects, such as the spatial dispersion of the NWP and the field-driven electron rescattering, which plays a minor role in the present study of light-induced molecular dissociation. The present WASP concept is general and molecule-independent and thus is applicable with minor changes to more complex molecules. Two typical examples of larger molecules, i.e., a multielectron diatomic molecule of $CO^+$ and an organic polyatomic molecule of $C_2H_2^{2+}$, are further examined and compared with the experiments (see Supplementary Materials [21] for details) to demonstrate the general applicability of the WASP approach in modeling the light-induced dynamics of various molecules.

## Methods

### Classical propagation of the NWP

The propagation of the NWP can be regarded as the classical motion of a particle with its reduced mass on the potential energy curves. The classical motion follows the Newtonian equation of motion driven by the force $F = -\mathrm{d}V(R)/\mathrm{d}R$ [28]. The motion of the NWP is initialized at the equilibrium distance of the neutral molecule with the initial kinetic energy matching the resonant dissociation condition, i.e., the NWP has enough kinetic energy to approach the internuclear distance of the resonant transition. Upon quantum-state transition, the kinetic energy and motion direction of the NWP is assumed to be unchanged on the respective potential energy curves. When the NWP moves with an internuclear distance exceeding 15 a.u., it will be regarded as an outgoing NWP denoting the dissociation of the molecule, and the kinetic energy the NWP carries is defined as the KER.

**Phase accumulation of the NWP**

When the NWP moves on the potential energy curves, a phase is attached to the NWP accumulated along the dissociation pathway. The accumulated phase of a specific dissociation pathway is described as follows [29]:

$$\varphi_J = -\sum_{k=1}^{n} E_k \Delta t_k + \sum_{k=1}^{n-1}(-1)^k \omega_k t_k + \sum_{k=1}^{n-1}(-1)^{k-1}\pi + \int_{R_0}^{R_\infty} p_J(R)\mathrm{d}R, \quad (6)$$

where the first term denotes the evolution phase of the nuclear eigenstate with the system energy $E_k$ and corresponding stretching time $\Delta t_k$, the second term denotes the evolution phase of the laser pulse with the absorption or emission of the photons of $\omega_k$, the third term denotes the phase associated with the symmetry change of the electronic wave function upon quantum-state transitions, and the last term denotes the dynamical phase of the NWP moving on the potential energy curves.

**Quantum-state transition description in molecular systems**

The description of the quantum-state transitions in a molecule is derived from the Rabi oscillation in atomic systems. For a two-level atom in a laser field, the population of the excited state demonstrates an oscillatory behavior [30]. When the photon energy equals the energy gap between the two states, i.e., $\hbar\omega = \Delta E$, the time-dependent population of the excited state can be derived under the rotating wave approximation as $P_e(t) = \sin^2(\Omega_0 t)$, where $\Omega_0 = D\mathcal{E}/2$ with $D$ the dipole between the two states and $\mathcal{E}$ the field strength of the laser field. While for a two-level molecular system, the nuclear motion greatly influences the hopping dynamics of the electron [25]. After the electron in the ground state is resonantly excited, the repulsive force between the two nuclei enables the molecular bond to stretch, and the energy gap of the two-level system varies. Thus, the oscillatory behavior of the population of the excited state with a fixed frequency would be interrupted, and the accumulated population transfer of the excited state can be approximated as $P_{1\omega} = \Omega_{1\omega}^2 \tau_{1\omega}^2$, where $\Omega_{1\omega} = D(R_{1\omega})\mathcal{E}/2$ is defined as the one-photon laser-molecule coupling at the resonant distance $R_{1\omega}$ and $\tau_{1\omega}$ denotes the effective duration of the one-photon transition in a molecule. Here the effective duration is calculated by the classical motion of the NWP within the distance region of $(R_{1\omega} - R_0, R_{1\omega} + R_0)$, i.e., $\tau_{1\omega} = t_{1\omega}(R_{1\omega} + R_0) - t_{1\omega}(R_{1\omega} - R_0)$, where $t_{1\omega}(R)$ denotes the bond-stretching time as a function of the internuclear distance and $R_0$ denotes the effective distance of the one-photon transition. The effective distance corresponds to the effective motion region where the detuning at the boundary is approximately equal to the laser-molecule coupling, i.e., $\Delta_0 = |\Delta E - \hbar\omega| \sim D\mathcal{E}_0 = 2\Omega_0$. With the corresponding detuning, the upper limit of the population of the excited state decreases to $P_e^{\text{upper}} = 4\Omega_0^2/(4\Omega_0^2 + \Delta_0^2) \sim 0.5$ so that the population of the excited state no longer increases. In other words, the effective distance defines an internuclear distance region where the laser-molecule coupling contributes to the whole transition amplitude. Thus, the one-photon transition amplitude in molecular systems is formulated as $A_{1\omega} = \sqrt{P_{1\omega}} = \Omega_{1\omega}\tau_{1\omega}$.

**Experimental technique**

We performed experiments using the phase-controlled linearly polarized two-color laser fields produced in a phase-locked Mach-Zehnder interferometer. The linearly polarized FW pulse (25 fs, 790 nm, 10 kHz) is generated from a multi-pass amplification Ti:sapphire laser system, and the linearly polarized SH pulse centered at 395 nm is produced by frequency doubling the FW pulse in a $\beta$-barium borate crystal.

The temporal overlap of the two pulses is controlled by a motorized delay stage in the FW arm. A phase-locking system based on the spatial interference of a reference continuum-wave laser at 532 nm is employed [31,32] to finely tune the two-color relative phase $\phi_L$. Further experimental details on the construction of the two-color laser fields can be found in Refs. [33-35]. The phase-controlled two-color laser pulse is tightly focused onto a supersonic gas jet of $H_2$ in an ultrahigh vacuum chamber of the cold-target recoil ion momentum spectroscopy reaction microscope [26,27]. The peak intensities of the FW and SH pulses in the interaction region can be adjusted by the neutral filters in the two arms, respectively. The two-color laser pulse first singly ionizes $H_2$ to $H_2^+$ which then dissociates into a neutral hydrogen atom and a proton. The positively charged proton is guided by a weak homogenous electric field and detected by a time- and position-sensitive microchannel plate detector at the end of the spectrometer. By measuring the time of flight and positions of the charged particle impacts, the three-dimensional momenta and kinetic energies of the protons are reconstructed event-by-event during the offline analysis.

**Asymmetry parameter**

For each dissociation pathway, the NWP eventually dissociates along either the $1s\sigma_g$ or $2p\sigma_u$ potential energy curve with opposite parities of the electron wavefunction. Hence, for the $H_2^+$ molecule placed horizontally, the electron localized at the left or right nucleus is determined by the coherent superposition of the NWPs on all gerade and ungerade states, i.e., $\psi_{\text{left}} = \sum \psi_g + \sum \psi_u$ and $\psi_{\text{right}} = \sum \psi_g - \sum \psi_u$. From the superimposed NWPs we can obtain the probabilities for measuring the electron localized at the left or right nucleus via $P_{\text{left}} = |\psi_{\text{left}}|^2$ and $P_{\text{right}} = |\psi_{\text{right}}|^2$. Therefore, the two-dimensional asymmetry distribution of the ejected proton can be calculated as [6,8]

$$\mathcal{A}(E_k, \theta_p, \phi_L) = \frac{P_{\text{left}}(E_k, \theta_p, \phi_L) - P_{\text{right}}(E_k, \theta_p, \phi_L)}{P_{\text{left}}(E_k, \theta_p, \phi_L) + P_{\text{right}}(E_k, \theta_p, \phi_L)}. \quad (7)$$

## Acknowledgments

This work was supported by the National Key R&D Program of China (Grants No. 2018YFA0306303 and No. 2018YFA0404802); the National Natural Science Foundation of China (Grants No. 11834004, No. 12227807, No. 11621404, No. 11925405, No. 91850203, No. 92150105, and No. 11904103); the Science and Technology Commission of Shanghai Municipality (Grant No. 21ZR1420100); S. Pan acknowledges the supports from the Academic Innovation Ability Enhancement Program for Excellent Doctoral Students of East China Normal University in 2021 (Grant No. 40600-30302-515100/141).

## Author contributions

S. Pan and J. Wu conceived the concept and initiated the study. S. Pan, P. Lu, X. Gong, R. Gong, W. Zhang, and L. Zhou performed the experiments and analyzed the data. S. Pan, Z. Zhang, and C. Hu performed the numerical simulations. All authors discussed the results and contributed to the writing of the manuscript. J. Wu, F. He, and H. Ni supervised and coordinated the work.

## Competing interests

The authors declare no competing interests.

## Data and materials availability

All data needed to evaluate the conclusions in the paper are present in the paper and/or the Supplementary Materials.

M. F. Kling, Steering proton migration in hydrocarbons using intense few-cycle laser fields, Phys. Rev. Lett. **116**, 193001 (2016).

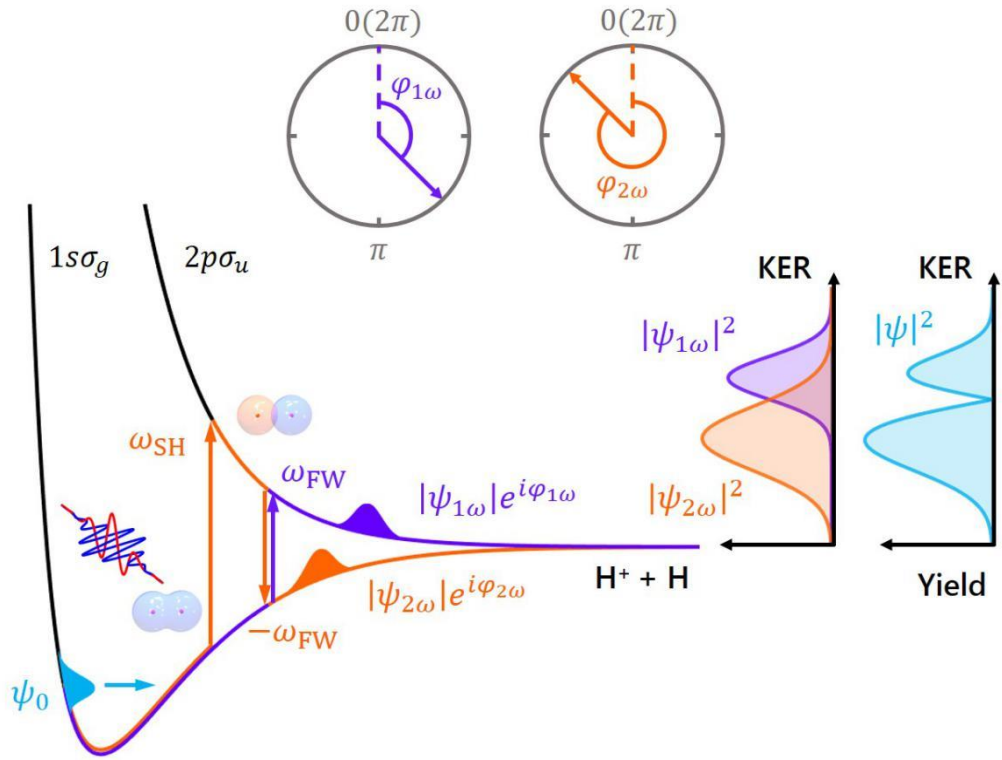

**Figure 1: Light-induced molecular dissociation.** Schematic illustration of light-induced dissociation dynamics of $H_2^+$ driven by an orthogonal two-color (OTC) laser field. Two black curves denote the potential energy curves of the electronic ground ($1s\sigma_g$) and excited ($2p\sigma_u$) states, respectively. The NWP on the ground state is launched by releasing one electron at the equilibrium internuclear distance of $H_2$, illustrated by the blue wave-packet. The created NWP subsequently propagates on these two potential curves, and here two pathways, for example, are shown with different colors ($1\omega_{SH} - 1\omega_{FW}$ pathway in orange and $1\omega_{FW}$ pathway in violet). Two dials with different colors denote the accumulated phases of the two pathways. The two panels on the right denote the KER spectra from the incoherent and coherent superposition of the two pathways, respectively.

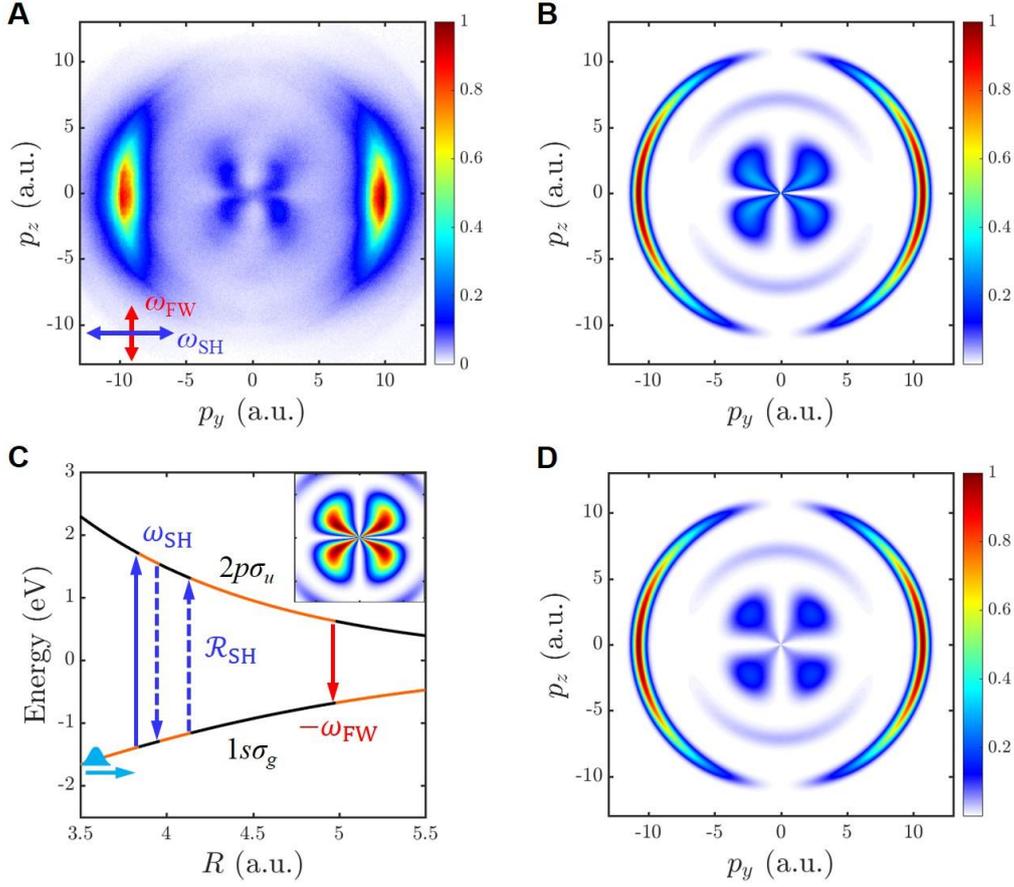

**Figure 2: Measured and simulated momentum distributions showing the proton anisotropy.** (**A**) Measured momentum distribution of ejected protons driven by the OTC pulse. The SH and FW pulses are polarized along the *y* axis and *z* axis, respectively. (**B**) Simulated momentum distribution of ejected protons driven by the OTC pulse. (**C**) Schematic illustration of the $1\omega_{\text{SH}} + 1\mathcal{R}_{\text{SH}} - 1\omega_{\text{FW}}$ pathway contributing to the production of the very low-energy protons of the butterfly structure shown as the inset. Here blue and solid red arrows denote the $1\omega_{\text{SH}}$ absorption and $1\omega_{\text{FW}}$ emission, respectively. Two blue dashed arrows denote the dynamical Rabi coupling driven by the SH field, termed as $\mathcal{R}_{\text{SH}}$, where the bound electron jumps between the two electronic states. (**D**) Simulated momentum distribution of protons driven by the OTC pulse without the dynamical Rabi coupling.

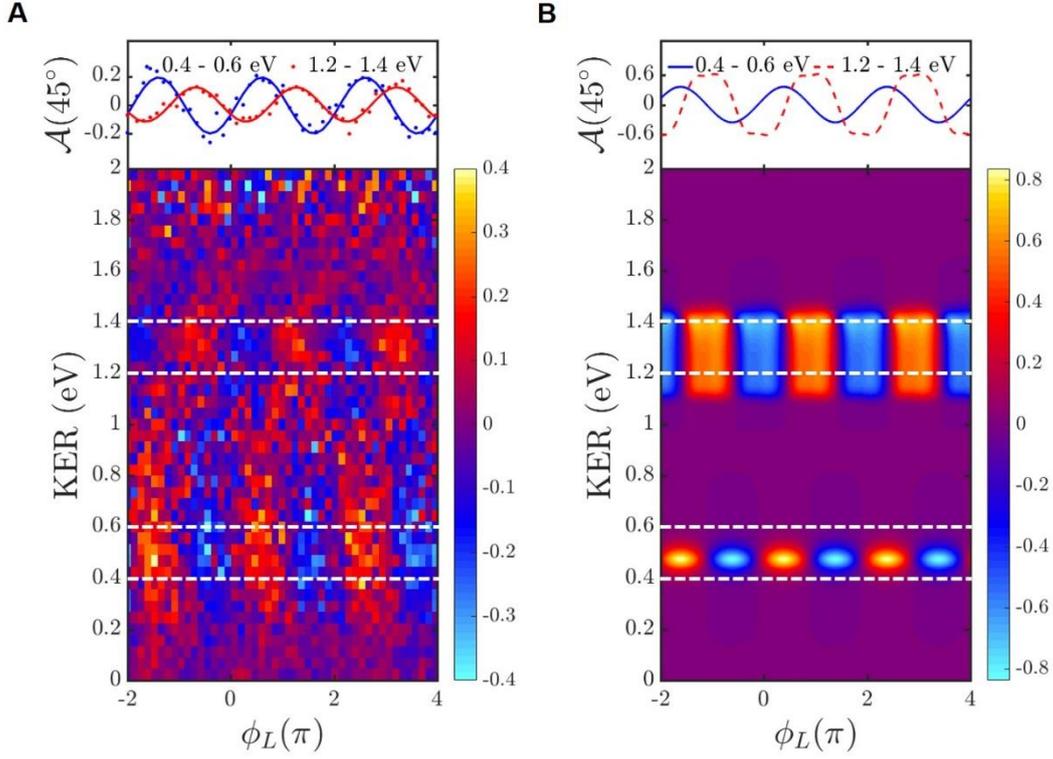

**Figure 3: KER-dependent asymmetric distributions.** (**A**) Measured two-dimensional maps of asymmetric distribution as a function of the relative phase $\phi_L$ of the OTC pulse and KER of the ejected protons within the range of $30° < \theta_p < 60°$. (**B**) Simulated two-dimensional map of the asymmetric distribution as in (**A**). Two KER regions with clear asymmetries are indicated between the white dashed lines, the asymmetries of which are plotted in the top panels.

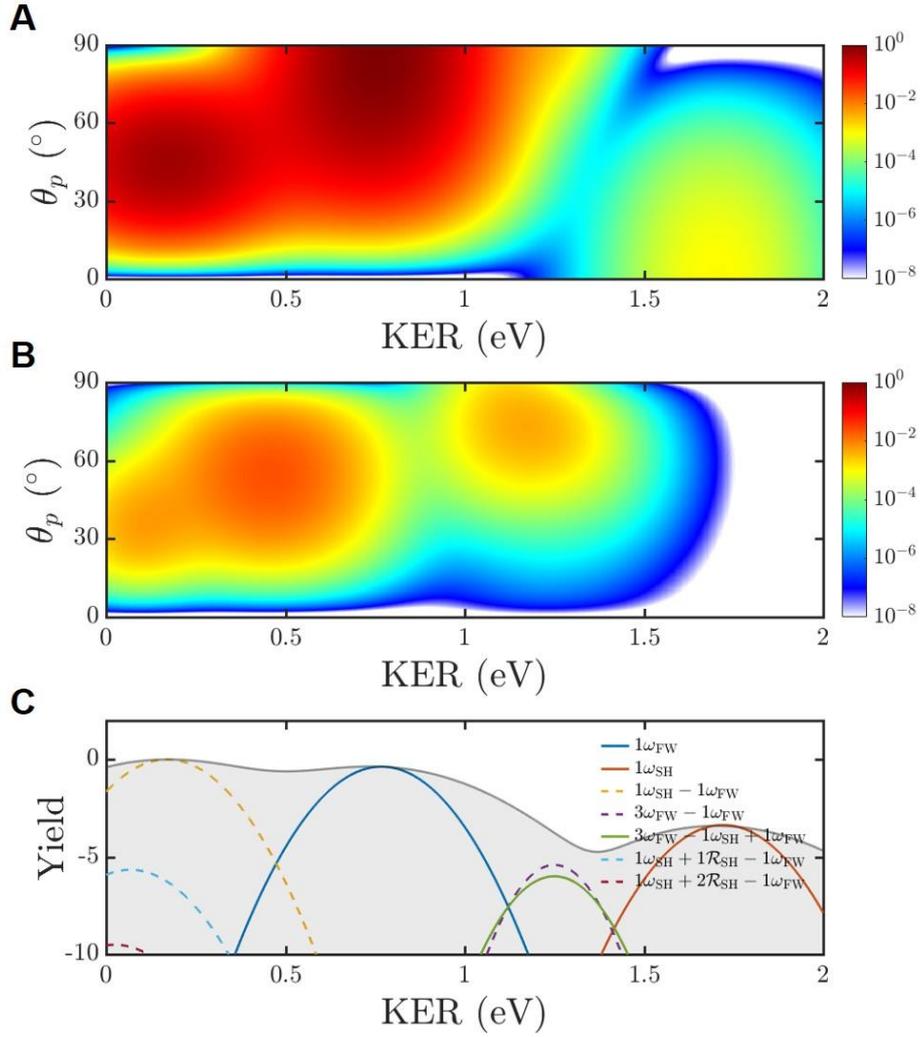

**Figure 4: Simulated KER spectra using the WASP model.** (**A**) Simulated two-dimensional KER spectra of protons versus the emission direction driven by an OTC pulse. (**B**) Differential KER spectra calculated by subtracting the KER spectra including the phase accumulation in (**A**) from the one without phase accumulation. (**C**) Simulated KER spectrum selecting the emission direction along 45° from (**A**). Solid and dashed curves denote the ungerade and gerade states where the NWP finally dissociates, respectively. Logarithm scales are used here.